\documentclass[aps,prc,twocolumn,superscriptaddress]{revtex4-2}
\usepackage{graphicx}% Include figure files
\usepackage{dcolumn}% Align table columns on decimal point
\usepackage{bm}% bold math
\begin{document}

% Use the \preprint command to place your local institutional report
% number in the upper righthand corner of the title page in preprint mode.
% Multiple \preprint commands are allowed.
% Use the 'preprintnumbers' class option to override journal defaults
% to display numbers if necessary
%\preprint{}

%Title of paper
\title{Search for triple and quadruple beta decay of $^{150}$Nd}

% repeat the \author .. \affiliation  etc. as needed
% \email, \thanks, \homepage, \altaffiliation all apply to the current
% author. Explanatory text should go in the []'s, actual e-mail
% address or url should go in the {}'s for \email and \homepage.
% Please use the appropriate macro foreach each type of information

% \affiliation command applies to all authors since the last
% \affiliation command. The \affiliation command should follow the
% other information
% \affiliation can be followed by \email, \homepage, \thanks as well.
%\author{A.S. Barabash$^1$, Ph. Hubert$^2$, A. Nachab$^2$, and V. Umatov$^1$ }
\author{A.S.~Barabash} 
\email{barabash@itep.ru}
\affiliation{NRC "Kurchatov Institute" - ITEP, 117218 Moscow, Russia}

\author{Ph.~Hubert}
\affiliation{CENBG, Universit\'e de Bordeaux, CNRS/IN2P3, F-33175 Gradignan, France}

\author{A.~Nachab}
\affiliation{D\'epartement de physique, Facult\'e Poly-disciplinaire de Safi, Universit\'e Cadi Ayyad, Route Sidi Bouzid BP 4162, 46000 Safi, Morocco }

\author{V.I.~Umatov}
\affiliation{NRC "Kurchatov Institute" - ITEP, 117218 Moscow, Russia}

%\email[]{Your e-mail address}
%\homepage[]{Your web page}
%\thanks{}
%\altaffiliation{}

%Collaboration name if desired (requires use of superscriptaddress
%option in \documentclass). \noaffiliation is required (may also be
%used with the \author command).
%\collaboration can be followed by \email, \homepage, \thanks as well.
%\collaboration{}
%\noaffiliation

\date{\today}

\begin{abstract}
% insert abstract here
Triple beta decay of $^{150}$Nd to the ground and excited states of $^{150}$Eu and quadruple beta decay of $^{150}$Nd to the excited states of $^{150}$Gd have been studied using a 400 cm$^3$ low-background HPGe detector and an external source consisting of 3046 g of natural 
Nd$_2$O$_3$ powder. A half-life limit for the quadruple beta decay to the 0$_1^+$ state of $^{150}$Gd was found to be $T_{1/2} (0\nu+4\nu) > 8.7\times10^{20}$ yr (90\% C.L.). For other (0$\nu$+4$\nu$) transitions to the 2$_1^+$, 3$_1^-$, 4$^+_1$, 2$_2^+$ and 2$^+_3$  excited states limits for the first time have been obtained at the level of ∼ $(6.1-9.5)\times10^{20}$ yr (90\% C.L.). We report here also the results for the first search for triple beta decay to the ground and excited final states of $^{150}$Eu. We find no evidence of this decay and set lower limits on the half-life in the range  $T_{1/2} (0\nu+3\nu) > (0.04-4.8)\times10^{20}$ yr (90\% C.L.).

\end{abstract}

% insert suggested PACS numbers in braces on next line
% insert suggested keywords - APS authors don't need to do this
%\keywords{}

%\maketitle must follow title, authors, abstract, \pacs, and \keywords
\maketitle

% body of paper here - Use proper section commands
% References should be done using the \cite, \ref, and \label commands
\section{Introduction}
Lepton-number-violating (LNV) processes could be directly linked to the possible Majorana nature of neutrinos. If Majorana mass terms are added to the standard model (SM) Lagrangian, processes appear that violate $\it L$ by two units ($\Delta L = 2$) \cite{SCH82}. Searches for $\Delta L = 2$  processes such as neutrinoless double beta decay ($0\nu2\beta$) have, therefore, been the goal of many experiments (see recent reviews \cite{VER16,BAR18,BAR19,DOL19}). The Majorana nature of the neutrino would also have interesting implications in many extensions of the Standard Model of particle physics. For instance the seesaw mechanism requires the existence of a Majorana neutrino to explain the lightness of neutrino masses \cite{MIN77,YAN80,GEL79,MOH80}. A Majorana neutrino would also provide a natural framework for lepton number violation, and particularly for the leptogenesis process which may explain the observed matter-antimatter asymmetry of the Universe \cite{FUK86}. However, it is most often overlooked that LNV and Majorana neutrinos are not necessarily connected. Models with $\Delta L = 3$ and $\Delta L = 4$ have some power in explaining naturally small Dirac masses of neutrinos and could mediate leptogenesis (see discussions in \cite{HEE13,HEE13a} and references therein). Processes with $\Delta L = 4$ could also be probed at the Large Hadron Collider (LHC), for example, in the pair production and decay of triplet-Higgs states to four identical charged leptons \cite{NEM17}.
In \cite{HEE13} the toy model with $\Delta L = 4$ was constructed which allows neutrinoless quadruple beta decay ($0\nu4\beta$):

\begin{equation}
(A,Z) \rightarrow (A,Z+4) + 4e^{-}
\end{equation}

This feature is discussed in more detail in \cite{DAS19}. 
Notice that the neutrino in the framework of this model is Dirac and the neutrinoless double beta decay is forbidden. The process (1) was discussed also in \cite{HIR18} and $\Delta L \ge 4$ lepton number violating processes were discussed in \cite{HIR18a}. 
%In \cite{HIR18} it was  argued  that if the null $0\nu2\beta$ decay signal is accompanied by a %$0\nu4\beta$ quadruple beta decay signal, then at least one neutrino should be a Dirac particle.
 Note also that quadruple decay with the emission of 4 neutrinos ($4\nu4\beta$) is not forbidden by any conservation low:

\begin{equation}
(A,Z) \rightarrow (A,Z+4) + 4e^{-} + 4\bar{\nu}
\end{equation}

This is simply a fourth order process for weak interaction. The authors of \cite{HEE13} pointed out three candidate nuclei for quadruple decay ($^{150}$Nd, $^{96}$Zr and $^{136}$Xe) and it was noted that for $^{150}$Nd the energy of 4$\beta$ transition is maximum (2084.2 keV \cite{WAN17}) and even transition to excited levels of the daughter nucleus $^{150}$Gd is possible. In this paper, we will also consider the triple beta decay of $^{150}$Nd (neutrinoless decay ($0\nu3\beta$) and decay with 3 neutrino emission ($3\nu3\beta$)): 

\begin{equation}
(A,Z) \rightarrow (A,Z+3) + 3e^{-}
\end{equation}

\begin{equation}
(A,Z) \rightarrow (A,Z+3) + 3e^{-} + 3\bar{\nu}
\end{equation}

In \cite{HEE13}, it was noted that neutrinoless triple beta decay is forbidden, since this process violates Lorentz invariance. This is true, but in this case the search for this process can serve as a test of Lorentz invariance. At the same time, the decay with the emission of three neutrinos is not forbidden by any conservation laws (this is a third-order process for the weak interaction). There are three candidates for this transition only. Besides $^{150}$Nd - $^{150}$Eu transition ($Q_{3\beta}$ =  1112.2 keV), it is also $^{48}$Ca - $^{48}$V ($Q_{3\beta}$ = 253.1 keV) and $^{96}$Zr - $^{96}$Tc ($Q_{3\beta}$ = 383.1 keV).

The search for 4$\beta$ decay was first carried out in the NEMO-3 experiment \cite{ARN17}. The $0\nu4\beta$ decay of $^{150}$Nd was investigated. For the transition to the ground state of the $^{150}$Gd the limit $T_{1/2} (0\nu) > (1.1-3.2)\times10^{21}$ yr (90\% C.L.) was obtained, depending on the model used for the kinematic distributions of the emitted electrons. 
%Notice, that the data of \cite{ARN17} can also be used to search for $0\nu3\beta$ decay. %Moreover, to establish the limit for $0\nu4\beta$ decay, a channel with three registered %electrons was investigated. 
Recently, first limit on the $0\nu4\beta$ decay of $^{150}$Nd to the 0$^+_1$ state (1207.13 keV) of $^{150}$Gd was obtained, $T_{1/2} (0\nu) > 1.76\times10^{20}$ yr (90\% C.L.) \cite{KID18}. Although the authors discuss only the limit for the $0\nu4\beta$ decay, it is clear that, in fact, this is also the limit for the $4\nu4\beta$ decay.

In this article, results of an experimental investigation of the 3$\beta(0\nu+3\nu)$ and 4$\beta(0\nu+4\nu)$  decays of $^{150}$Nd to the ground and excited states of $^{150}$Eu  and excited states of $^{150}$Gd are presented. Mass-chain decay scheme is shown in Fig. 1 and demonstrates the fact that $^{150}$Nd energetically can decay to $^{150}$Sm (double beta decay), $^{150}$Eu (triple beta decay) and $^{150}$Gd (quadruple beta decay). The decay schemes of $^{150}$Nd to $^{150}$Gd and $^{150}$Eu are presented in Fig. 2 and Fig. 3. The simplified decay scheme of $^{150}$Eu and $^{150m}$Eu are presented in Fig. 4 and Fig. 5, respectively. The measurements have been carried out using a HPGe detector to look for $\gamma$ -ray lines corresponding to the decay scheme. We used experimental data obtained in \cite{BAR04,BAR09}, that previously used to investigate  $2\beta(0\nu+2\nu)$ decay of $^{150}$Nd to the excited states of $^{150}$Sm.

%\begin{figure*}
%\includegraphics{nd_gd}% Here is how to import EPS art
%\caption{\label{fig:wide}Use the figure* environment to get a wide
%figure that spans the page in \texttt{twocolumn} formatting.}
%\end{figure*} 

\begin{figure*}[htb]
\begin{center}
\includegraphics[width=0.7\textwidth]{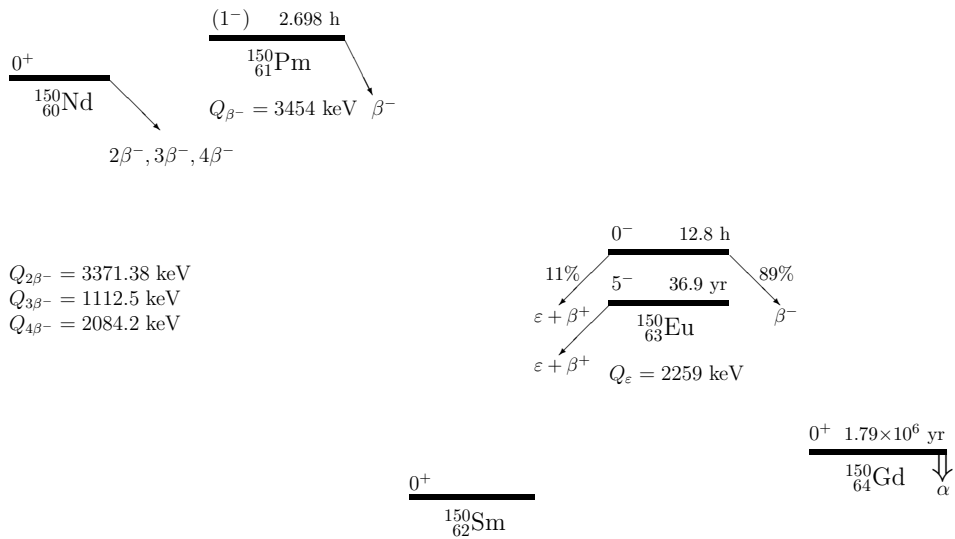}
%\vspace{-1.8truecm.}
\caption{Mass-chain decay scheme (taken from \cite{BAS13}).}
\label{fig:1}
\end{center}
\end{figure*} 
 
\begin{figure*}[htb]
\begin{center}
\includegraphics[width=0.6\textwidth]{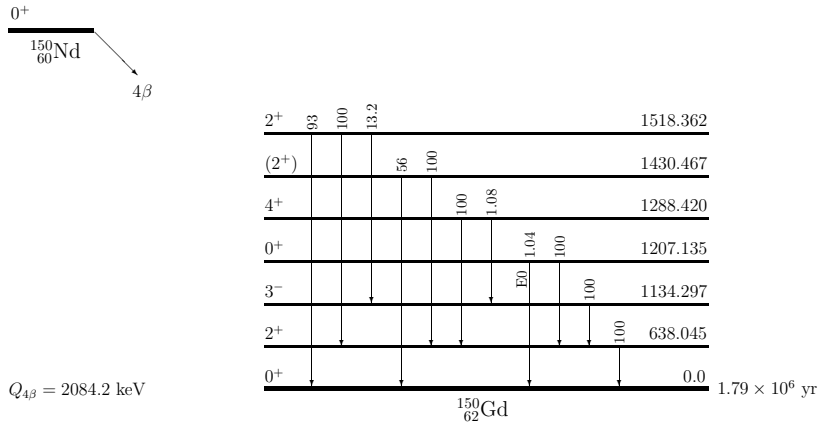}
%\vspace{-1.8truecm.}
\caption{Level scheme of $^{150}$Nd 4$\beta$ decay to $^{150}$Gd (taken from \cite{BAS13}). Energy levels are in keV. Relative branching ratios from each level are presented.}
\label{fig:2}
\end{center}
\end{figure*}

\begin{figure*}[htb]
\begin{center}
\includegraphics[width=0.6\textwidth]{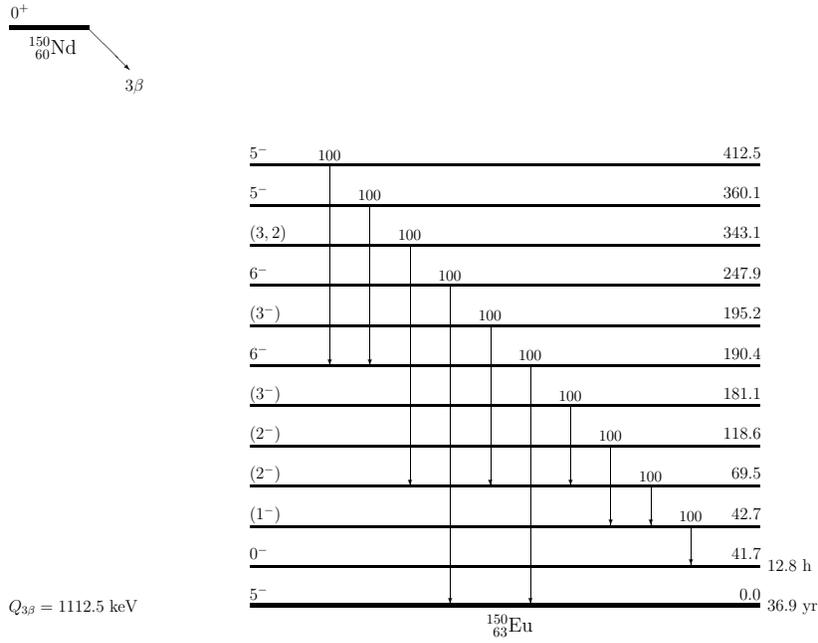}
%\vspace{-1.8truecm.}
\caption{Level scheme of $^{150}$Nd 3$\beta$ decay to $^{150}$Eu (taken from \cite{BAS13}). The diagram shows the levels corresponding to transitions with the most intense $\gamma$-lines. Energy levels are in keV. Relative branching ratios from each level are presented.}
\label{fig:3}
\end{center}
\end{figure*}

\begin{figure*}[htb]
\begin{center}
\includegraphics[width=0.6\textwidth]{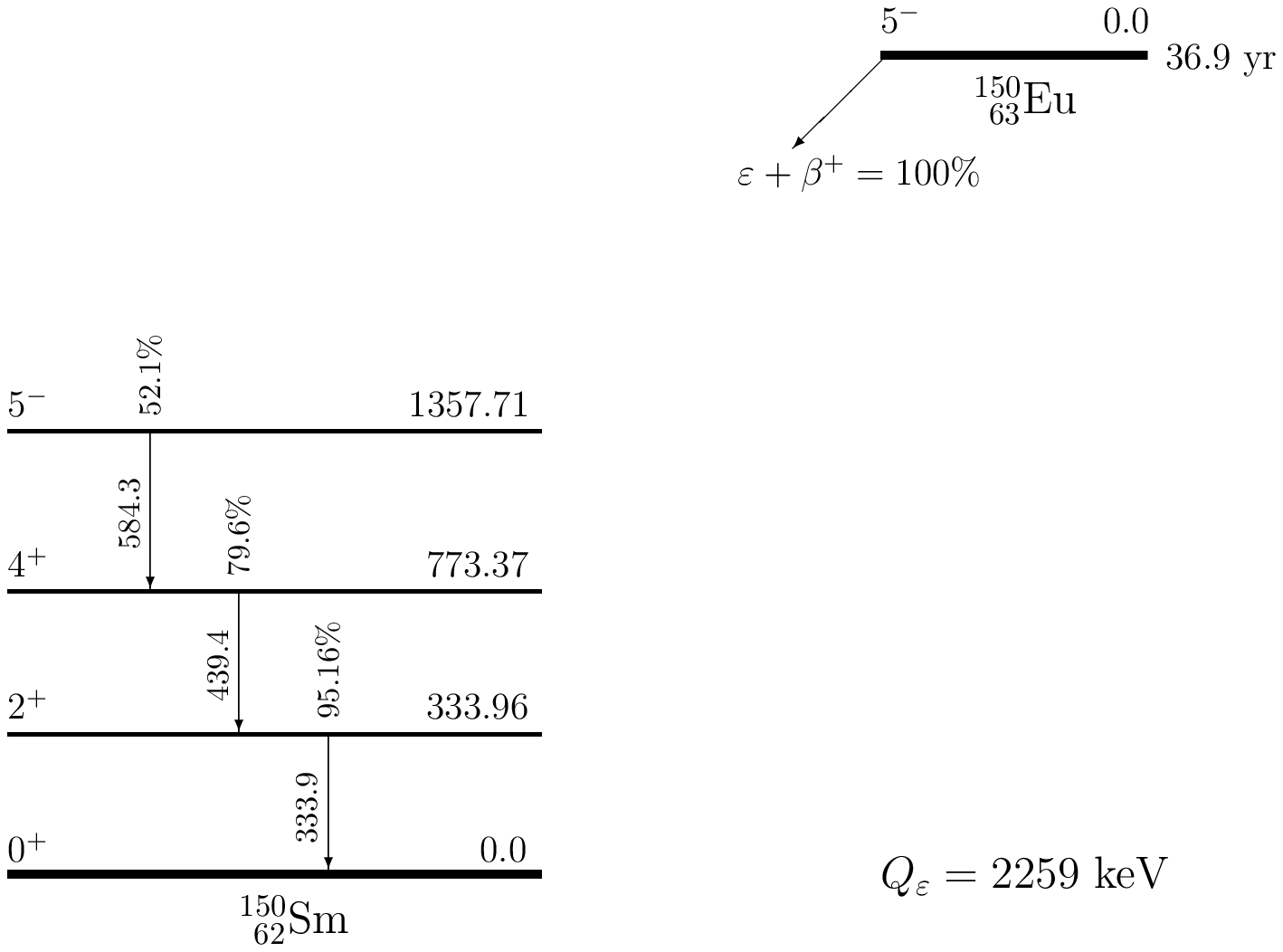}
%\vspace{-1.8truecm.}
\caption{Simplified decay scheme of $^{150}$Eu  (taken from \cite{BAS13}). Energy levels and energy of $\gamma$-rays are in keV. The intensity of gamma lines is given as a percentage of parent decays.}
\label{fig:4}
\end{center}
\end{figure*}

\begin{figure*}[htb]
\begin{center}
\includegraphics[width=0.6\textwidth]{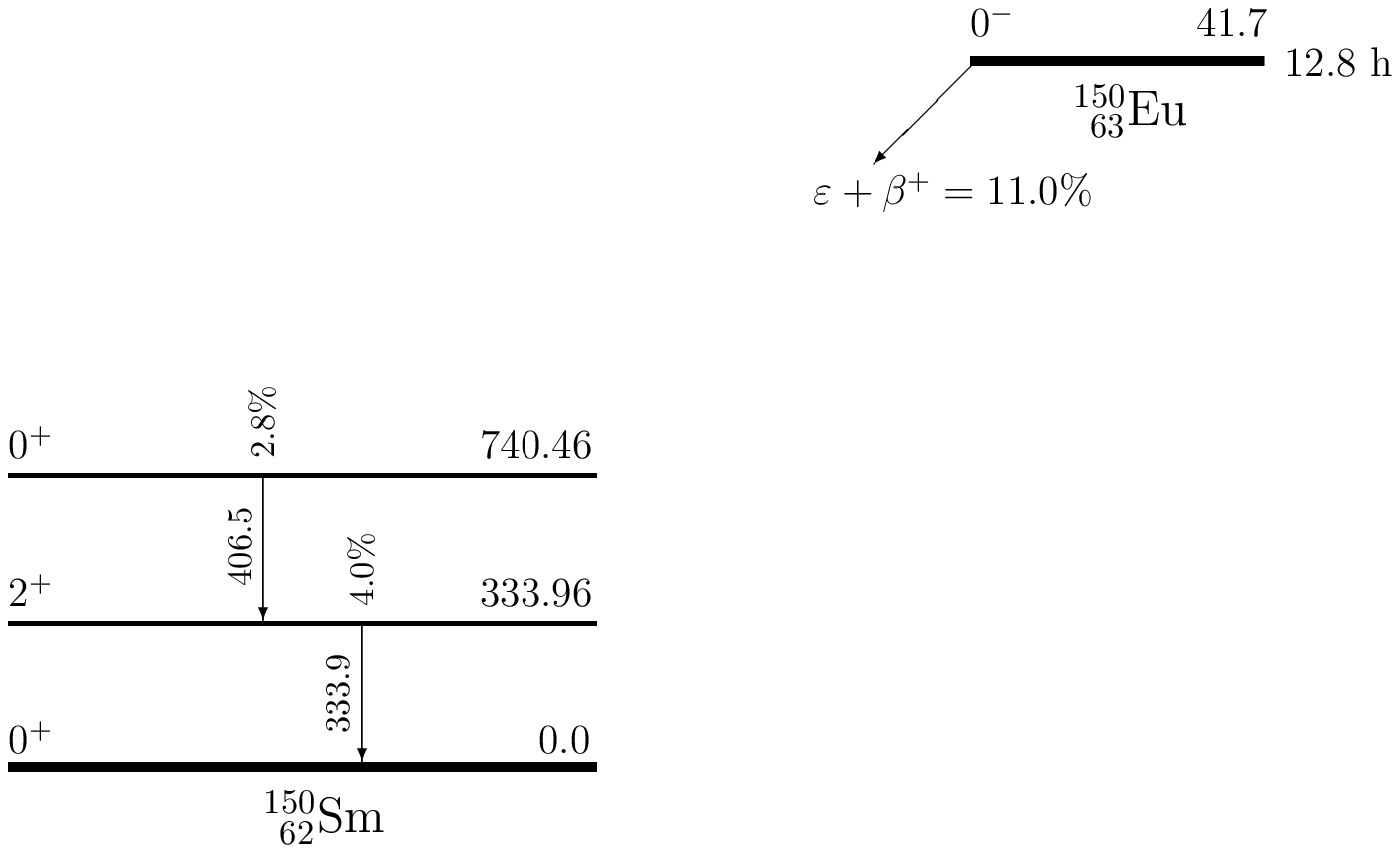}
%\vspace{-1.8truecm.}
\caption{Simplified decay scheme of metastable state of $^{150}$Eu (taken from \cite{BAS13}). Energy levels and energy of $\gamma$-rays are in keV. The intensity of gamma lines is given as a percentage of parent decays.}
\label{fig:5}
\end{center}
\end{figure*}
 
\section{Experiment}

The experimental work was performed in the Modane Underground Laboratory (depth of 4800 m w.e.). A 400 cm$^3$ low-background HPGe detector was used to measure a 3046 g sample of Nd$_2$O$_3$ powder in a special Marinelli delrin box which was placed on the detector endcap.
Taking into account the natural abundance of $^{150}$Nd (5.64\%) there are 153 g of $^{150}$Nd (or $6.14\times10^{23}$ nuclei of $^{150}$Nd) in the sample. Data was collected for 11320.5 hours.

The Ge spectrometer is composed of a p-type, 400 cm$^3$ crystal. The cryostat, endcap and the other mechanical parts have been made of a very pure Al-Si alloy. 
%The cryostat has a J-type geometry to shield the crystal from radioactive impurities in the %dewar. 
The passive shielding consisted of 3-10 cm of OFHC copper inside 15 cm of ordinary lead. To remove $^{222}$Rn gas, one of the main sources of the background, a special effort was made to minimize the free space near the detector. In addition, the passive shielding was enclosed in an aluminum box flushed with high-purity nitrogen.

The electronics consisted of spectrometric amplifiers and a 8192 channel ADC. The energy calibration was adjusted to cover the energy range from 50 keV to 3.5 MeV, and the energy resolution was 2.0 keV for the 1332-keV line of $^{60}$Co. The electronics were stable during the experiment due to the constant conditions in the laboratory (temperature  of $\approx 23^\circ$ C, hygrometric degree of $\approx 50$\%).  A daily check on the apparatus assured that the counting rate was statistically constant. 

The current data of accepted values for different isotopes published in Nuclear Data Sheets \cite{BAS13} were used for analysis of the energy spectrum. The photon detection efficiency for each investigated process has been calculated with the CERN Monte Carlo code GEANT 3.21. Special calibration
measurements with radioactive sources and powders containing well-known $^{226}$Ra activities confirmed that the accuracy of these efficiencies is about 10\%.

The dominate detector backgrounds come from natural $^{40}$K, radioactive chains of $^{232}$Th and $^{235,238}$U, man-made and/or cosmogenic activities of $^{137}$Cs and $^{60}$Co. The sample was found to have a large activity of $^{40}$K (46.3 mBq/kg).
Additionally long-lived radioactive impurities were observed in the sample, but with much weaker activities, i.e.  $^{137}$Cs (0.089 mBq),  $^{176}$Lu (0.450 mBq/kg),  $^{138}$La (0.068 mBq/kg),  $^{133}$Ba (0.155 mBq/kg), etc. 
%In our case the most important isotopes  contributing to energy ranges of the investigated %transitions are $^{214}$Bi (1.15 mBq/kg), $^{228}$Ac (0.93 mBq/kg), $^{227}$Ac (0.62 mBq/kg) %and their daughters.

Figures 6, 7, 8 and 9 show the energy spectrum in the ranges of interest.

\begin{figure*}
\includegraphics[width=0.6\textwidth]{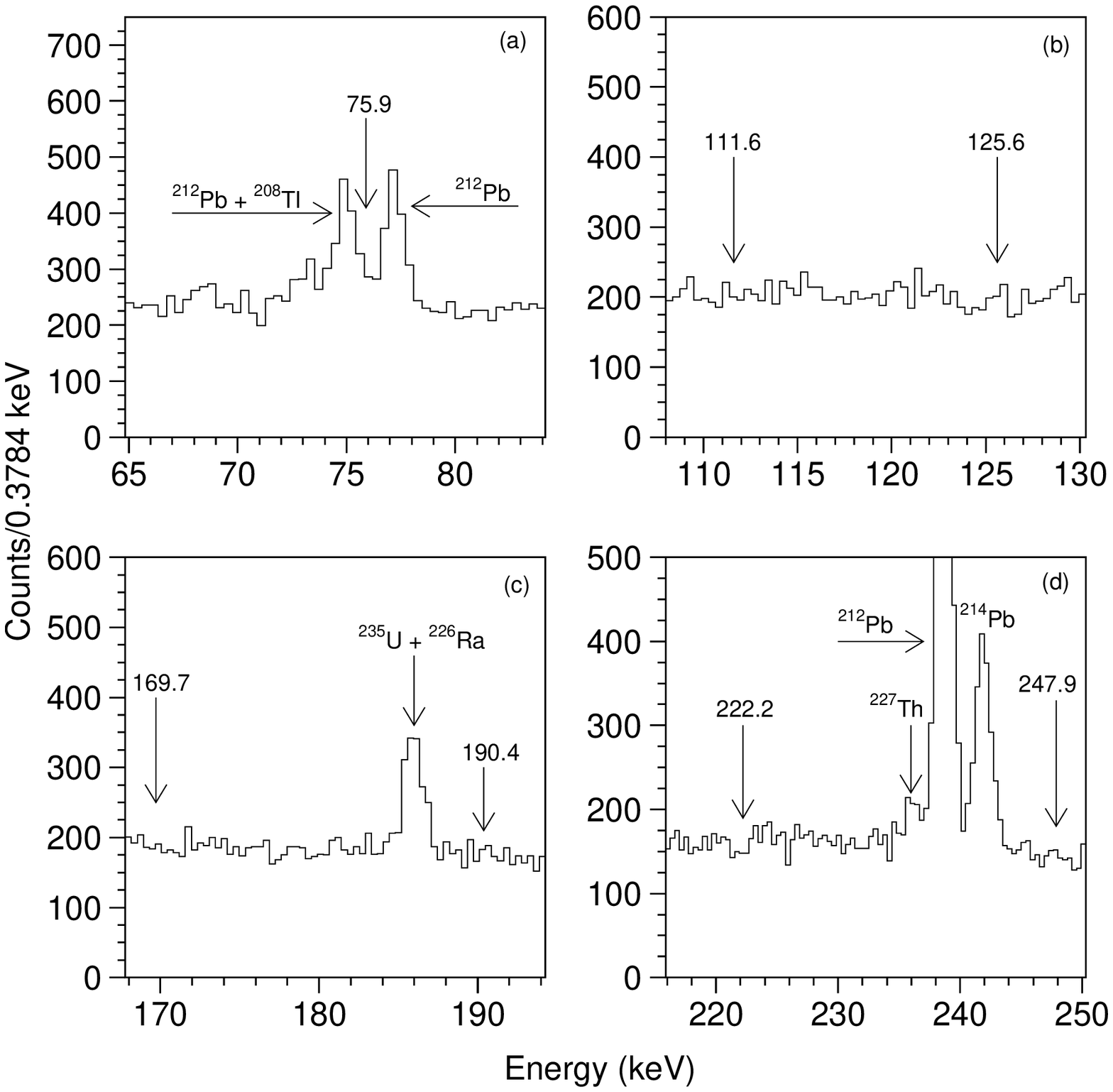}% Here is how to import EPS art
\caption{\label{fig:6}Energy spectrum with natural Nd$_2$O$_3$ powder in the ranges of 
investigated $\gamma$-rays: [65-84] keV (a), [108-130] keV (b), [168-192] keV (c) and 
[216-250] keV (d).}
\end{figure*}
 
\begin{figure*}
\includegraphics[width=0.6\textwidth]{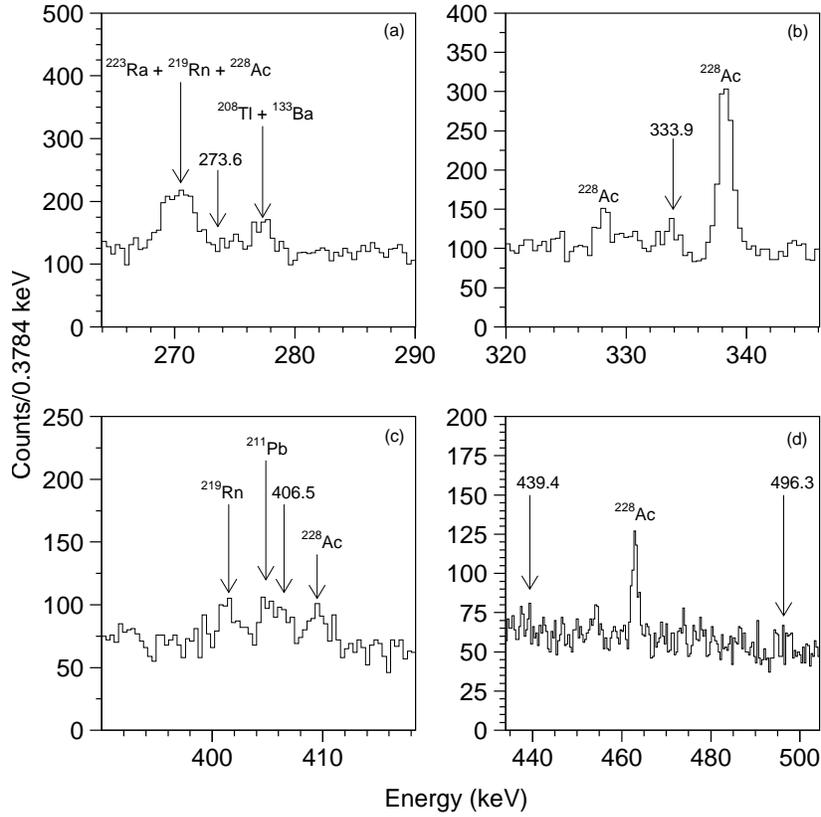}% Here is how to import EPS art
\caption{\label{fig:7}Energy spectrum with natural Nd$_2$O$_3$ powder in the ranges of 
investigated $\gamma$-rays: [264-290] keV (a), [320-343] keV (b), [390-414] keV (c) and [435-505] keV (d).}
\end{figure*}

\begin{figure*}
\includegraphics[width=0.6\textwidth]{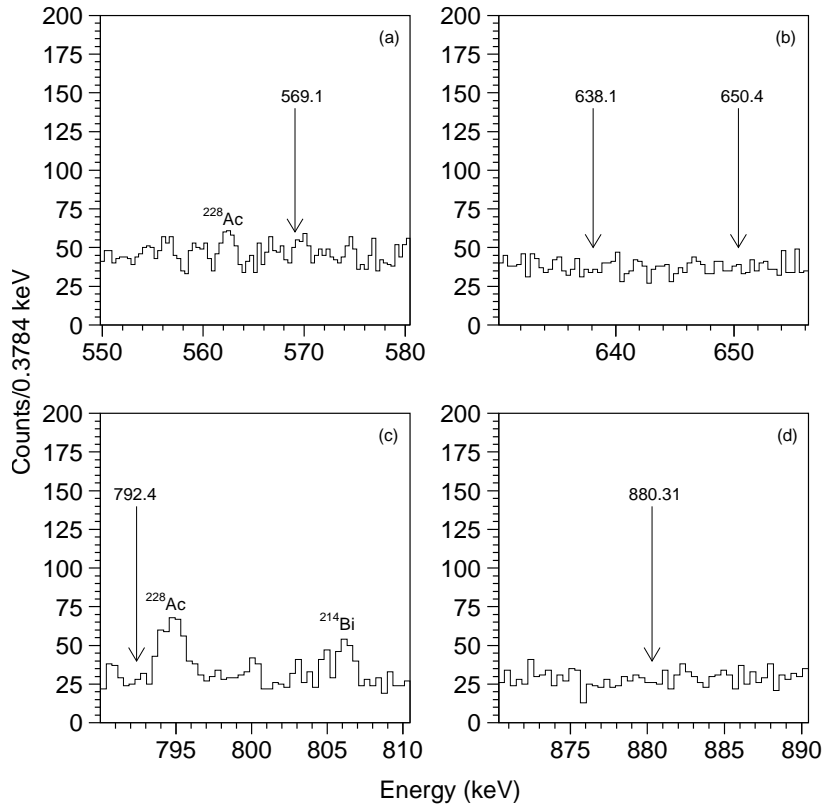}% Here is how to import EPS art
\caption{\label{fig:8}Energy spectrum with natural Nd$_2$O$_3$ powder in the ranges of 
investigated $\gamma$-rays: [550-580] keV (a), [630-656] keV (b), [790-810] keV (c) and (870-890) keV (d).}
\end{figure*}

\begin{figure*}
\includegraphics[width=0.6\textwidth]{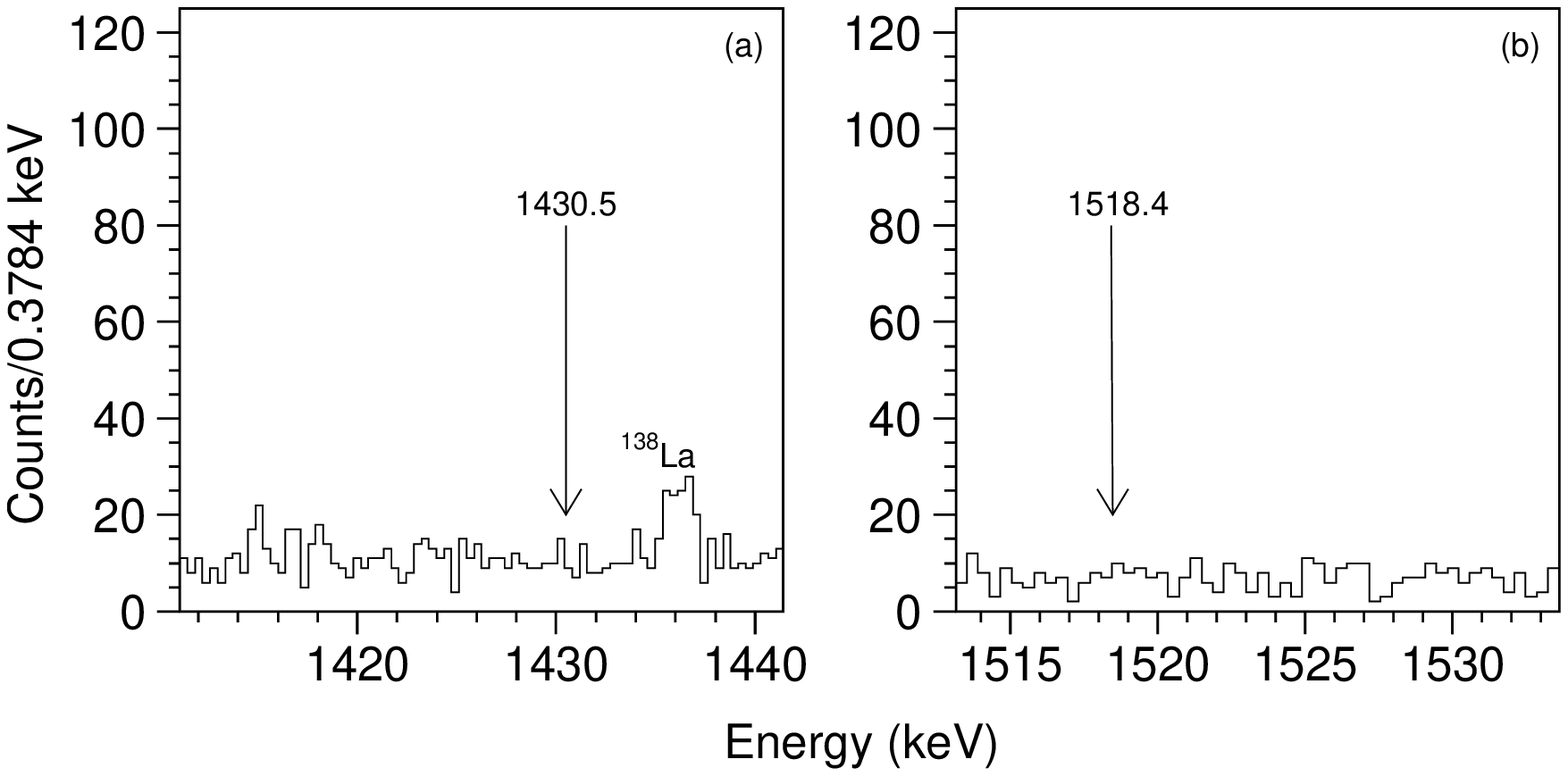}% Here is how to import EPS art
\caption{\label{fig:9}Energy spectrum with natural Nd$_2$O$_3$ powder in the ranges of 
investigated $\gamma$-rays: [1410-1440] keV (a) and [1513-1534] keV (b).}
\end{figure*}

%\begin{figure}[h!]
%\begin{center}
%\includegraphics[width=10cm]{fig2}% This is a *.eps file
%\end{center}
%\caption{Predictions on $\langle m_{\nu} \rangle$ from neutrino oscillations versus the %lightest neutrino mass m$_0$ in cases of NO (the blue region) and IO (the red region). The 2$%\sigma$ and 3$\sigma$ values of neutrino oscillation parameters are considered \cite{CAP17}.  %The excluded region by cosmological data m$_0$ is  presented in yellow ($>$ 30 meV for the NO %and $>$ 16 meV for the IO). The value of $\Sigma m_{\nu}$ $<$ 0.12 eV has been used %\cite{PLANCK}.}\label{fig:2}
%\end{figure}

%\begin{figure}[htb]
%\begin{center}
%\includegraphics[width=0.35\textwidth]{fig3.eps}
%\vspace{-1.8truecm.}
%\caption{Theoretical distributions of the individual electron energies for three models of %$^{100}$Mo $2\nu\beta\beta$ decay: HSD, SSD and SSD-3.%
%}
%\label{fig:3}
%\end{center}
%\end{figure}

\section{Analysis and results} 

\subsection{Search for 4$\beta(0\nu+4\nu)$  processes in $^{150}$Nd}

Quadruple beta decays of $^{150}$Nd to 2$^+_1$ (638.045 keV), 3$^-_1$ (1134.297 keV), 0$^+_1$ (1207.135 keV), 4$^+_1$ (1288.420 keV), 2$^+_2$ (1430.467 keV) and 2$^+_3$ (1518.362 keV) excited states of $^{150}$Gd have been investigated. 

\subsubsection{Decay to the 2$^+_1$ excited state}
To search for this transition one has to look for a $\gamma$-ray with an energy 
of 638.05 keV. The detection efficiency is 2.23\%. As one can see from Fig. 8, there is no statistically significant peak at this energy. So one can only give 
the lower half-life limit on the transition to the 2$^+_1$ excited state
of $^{150}$Gd, $T_{1/2} > 7.5\times 10^{20}$ yr. The limit has been calculated using the likelihood function described in  
\cite{BAR96,BAR96a} which takes into account all the peaks identified above 
as background.
This result is presented in Table 1.

\begin{table*}
\caption{\label{tab:table1}Experimental limits for 4$\beta(0\nu+4\nu)$ decay 
of $^{150}$Nd to the excited states of $^{150}$Gd. All limits are given at the 90\% C.L.}
\begin{ruledtabular}
\begin{tabular}{cccc}
 & &\multicolumn{2}{c}{$T_{1/2}^{0\nu+4\nu}, (\times10^{20} yr)$}\\
 Excited state, keV & Energy of $\gamma$-rays, keV (efficiency)
& This work & Ref. \cite{KID18}\\ \hline
 2$^+_1$ (638.05) & 638.05 (2.23\%) & 7.52 & - \\
 3$^-_1$ (1134.30) & 496.25 (2.27\%)+638.05 (2.09\%)) & 7.79 & - \\
 0$^+_1$ (1207.14) & 569.09 (2.11\%)+638.05 (2.06\%)) & 8.70 & 1.76 \\
 4$^+_1$ (1288.42) & 638.05 (2.05\%)+650.37 (2.02\%)) & 8.98 & - \\
 2$^+_2$ (1430.47) & 638.05 (1.33\%)+792.42 (1.20\%))+1430.47 (0.62\%) & 9.49 & - \\
 2$^+_3$ (1518.36) & 638.05 (1.13\%)+880.31 (0.86\%))+1518.36 (0.73\%) & 6.10 & - \\
\end{tabular}
\end{ruledtabular}
\end{table*}

\subsubsection{Decay to the 0$^+_1$ excited state}
The transition is accompanied by two $\gamma$-rays with energies of 
569.09 keV and 638.05 keV. The detection photopeak efficiencies are equal to 2.11\% 
at 569.09 keV and 2.06\% at 638.05 keV.
Fig. 8 shows the energy spectrum in the ranges of interest. There are no statistically significant 
peaks at
these energies. Using the same technique as above the lower 
half-life limits  $8.7\times 10^{20}$ yr are found for the 
transitions (Table 1). Table 1 also presents other valuable data on 
this transition.

\subsubsection{Decay to the 3$^-_1$, 4$^+_1$, 2$^+_2$ and 2$^+_3$ excited states}

To search for these transitions one has to look for $\gamma$-rays with 
energies of 496.25, 638.05, 650.37, 792.42, 880.31, 1430.47 and 1518.36 keV.  
Figures 7-9 show that there are no statistically significant 
peaks at
these energies. Using the same technique as above the lower 
half-life limits are found within $(6.1-9.5) \cdot 10^{20}$ yr for the 
transitions (see Table 1).

\subsection{Search for 3$\beta(0\nu+3\nu)$  processes in $^{150}$Nd}

Triple beta decays of $^{150}$Nd to the ground state of $^{150}$Eu ($5^-$), metastable state of $^{150}$Eu ($0^-_1$; 41.7 keV), 2$^-_2$ (118.6 keV), 3$^-_1$ (181.1 keV), 6$^-_1$ (190.4 keV), 3$^-_2$ (195.2 keV), 6$^-_2$ (247.9 keV), (3,2) (343.1 keV), 5$^-_1$ (360.1 keV) and 5$^-_2$ (412.5 keV)  excited states of $^{150}$Eu have been investigated. 

\subsubsection{Decay to the ground state of $^{150}$Eu}

In this case, the $^{150}$Nd (0$^+$) enters the ground state of the $^{150}$Eu (5$^-$). This nucleus is unstable (half-life is 36 years) and turns to $^{150}$Sm by electron capture. Transitions go to the excited states of $^{150}$Sm, which is accompanied by the emission of cascades of gamma quanta. 
%Using these gamma rays we search for 3$\beta(0\nu+3\nu)$ decays of $^{150}$Nd. 
In this case, we used the 439.40 keV line, which accompanies 79.6\% of the $^{150}$Eu decays. The registration efficiency of these gamma quanta is 2.5\%. But to calculate detection efficiency full decay scheme was used. 
The experiment used a Nd$_2$O$_3$ sample, produced 15 years before the measurements. It means that $^{150}$Eu from the possible decay of $^{150}$Nd could accumulate in the sample all this time. This was taken into account when determining the limit on the possible transition $^{150}$Nd-$^{150}$Eu. The resulting limit $0.5\times10^{20}$ yr is shown in Table 2.

\begin{table*}
\caption{\label{tab:table2}Experimental limits for 3$\beta(0\nu+3\nu)$ decay 
of $^{150}$Nd to the ground and excited states of $^{150}$Eu. All limits are given at the 90\% C.L.}

\begin{ruledtabular}
\begin{tabular}{cccc}
 & &\multicolumn{2}{c}{$T_{1/2}^{0\nu+3\nu}, (\times10^{20} yr)$}\\
 Excited state, keV & Energy of $\gamma$-rays, keV (efficiency)
& This work & Previous work\\ \hline
 5$^-_{g.s.}$, 36.9 yr & 439.40 (1.71\%) & 0.50 & - \\
 0$^-_1$ (41.7 keV), 12.8 h & 333.96 (0.10\%) & 0.04 & - \\
 2$^-_1$ (118.6 keV) & 75.9 (0.12\%) & 0.30 & - \\
 3$^-_1$ (181.1 keV) & 111.6 (0.48\%) & 0.63 & - \\
 6$^-_1$ (190.37 keV) & 190.4 (1.52\%) & 1.81 & - \\
 3$^-_2$ (195.2 keV) & 125.6 (0.66\%) & 0.78 & - \\
 6$^-_2$ (247.89 keV) & 247.9 (2.24\%) & 3.27 & - \\
 (3,2) (343.1 keV) & 273.6 (2.11\%) & 3.16 & - \\
 5$^-_1$ (360.14 keV) & 169.7 (1.28\%)+190.4 (1.44\%) & 2.24 & - \\
 5$^-_2$ (412.53 keV) & 190.4 (1.39\%)+222.2 (1.86\%) & 4.83 & - \\
 
\end{tabular}
\end{ruledtabular}
\end{table*}

\subsubsection{Decay to the 0$^-_1$ metastable state of $^{150}$Eu}

In this case, the $^{150}$Nd (0$^+$) enters the metastable state of the $^{150}$Eu (0$^-$). This nucleus is unstable (half-life is 12.6 hours) and disintegrates through beta decay (89\%) and electron capture (11\%). In electron capture, in 4\% of cases, a transition to the excited states of $^{150}$Sm occurs, and in 100\% of cases this leads to the emission of a gamma quantum with an energy of 333.9 keV. To search for this transition one has to look for a $\gamma$-ray with an energy of 333.9 keV. The final detection efficiency is 0.10\%. The analysis given in \cite{BAR09} shows that the excess of events at 333.9 keV is mainly due to the double beta decay of $^{150}$Nd to the 0$^+_1$ excited state of $^{150}$Sm. So one can only give the lower half-life limit on the metastable state of the $^{150}$Eu (0$^-$). The limit has been calculated using the likelihood function described in Refs. \cite{BAR96,BAR96a} which takes into account the peak identified above as background. The resulting limit $4\times10^{18}$ yr has been obtained and is shown in Table 2.

\subsubsection{Decay to the 2$^-_2$ (118.6 keV), 3$^-_1$ (181.1 keV), 6$^-_1$ (190.4 keV), 3$^-_2$ (195.2 keV), 6$^-_2$ (247.9 keV), (3,2) (343.1 keV), 5$^-_1$ (360.1 keV) and 5$^-_2$ (412.5 keV) excited states}

To search for these transitions one has to look for $\gamma$-rays with 
energies of 75.9, 111.6, 190.4, 125.6, 247.9, 273.6, 169.7 and 222.2 keV.  
As one can see from figures 6 and 7, there are no statistically significant 
peaks at
these energies. Using the same technique as above \cite{BAR96,BAR96a} the lower 
half-life limits are found within $(0.04-4.83) \cdot 10^{20}$ y for the 
transitions (see Table 2).

\section{Discussion and Conclusion}

The results are shown in Tables 1 and 2. Note that the limit on 4$\beta(0\nu+4\nu)$ decay of $^{150}$Nd to the 0$_1^+$ excited state of $^{150}$Gd obtained in this work is five times stronger than the limit obtained in  \cite{KID18}. For other (0$\nu$+4$\nu$) transitions to the 2$_1^+$, 3$_1^-$, 4$^+_1$, 2$_2^+$ and 2$^+_3$  excited states limits for the first time have been obtained. Results for 3$\beta(0\nu+3\nu)$ decay are obtained for the first time. Unfortunately, there are no "real" (accurate) theoretical calculations for these processes. Therefore, we cannot compare our results with theoretical predictions. It is clear only that such transitions should be strongly suppressed compared to the $2\beta$ decay processes. However, as indicated in \cite{HEE13,DAS19}, under certain conditions it is possible to have significant enhancement for the $4\beta (0\nu)$ decay process. In \cite{HIR18a}  it was demonstrated (using "qualitative" estimates) that, under the most favorable assumptions, the half-life of $^{150}$Nd for $0\nu4\beta$ decay is 
$\sim 10^{41}$ yr  \cite{HIR18a} (this can be compared with the best present experimental limit, $T_{1/2} > (1.1-3.2)\times10^{21}$ yr \cite{ARN17}). Thus, it is clear that sensitivity of present experiments is very far from theoretical expectations. And, apparently, the prospects for the search for $0\nu3\beta$ decays look even more pessimistic. It seems that the search for processes with the emission of neutrinos is more likely to succeed. Unfortunately for these processes there are not even qualitative theoretical estimates.
%Therefore, both theoretical studies and experimental verification of the possibility of observing such %processes are necessary. 

In addition to direct counter experiments, other types of experiments can be used. 
In work \cite{HEE13}, a radiochemical experiment is proposed to search for $4\beta$ decay of $^{150}$Nd, given the fact that the daughter nucleus ($^{150}$Gd) is radioactive (alpha decay, $T_{1/2} = 2\times10^6$ yr). In such type of experiment, $2\beta$ decay of $^{238}$U \cite{TUR91} was recorded, for example. In principle, one can offer here a geochemical experiment too, by analogy with geochemical experiments with $^{130}$Ba \cite{MES01} , $^{100}$Mo \cite{HID04} or $^{96}$Zr \cite{KAW93}.
In the case of $3\beta$ decay, radioactive nuclei $^{150}$Eu (5$^-$) and $^{150m}$Eu (0$^-$) are formed, with a lifetime of 36 yr and 12.6 h, respectively. This circumstance can be used in the conducting of radiochemical experiments too. The sensitivity of such experiments can be up to $\sim 10^{25}-10^{27}$ yr (the estimation is made using the parameters of radiochemical experiments on the registration of solar neutrinos, where the mass of the analyte can be 50 tons \cite{GAV11}). Of course, radiochemical and geochemical experiments are sensitive to complete decay and cannot distinguish neutrinoless decay from decay with emission of neutrinos. It is unknown a priori what type of decay is more likely, neutrinoless decay or decay with emission of neutrinos. It is possible that decay with emission of neutrinos will be more probable and we have more chance to see this decay in future experiments.
Theoretical studies are needed on this issue. But, in any case, the observation of such transitions will be extremely important and interesting. 

\begin{acknowledgments}

Part of this work was supported by Russian Science Foundation (grant No. 18-12-00003).

\end{acknowledgments}

\bibliography{basename of .bib file}

\end{document}